\documentclass[floatfix,twocolumn,aps,prb,superscriptaddress]{revtex4-2}
\usepackage[T1]{fontenc}
\usepackage{float}
\usepackage{multirow}
\usepackage{amsmath}
\usepackage{graphicx}
\usepackage{xcolor}
\usepackage{hyperref}

\begin{abstract}
Bond reconstruction in vacancy-related structures affects their formation energies, symmetries, and electronic and optical properties. Using density functional theory, we investigate bond-reconstruction mechanisms in monovacancies and vacancy aggregates in monolayer silicon carbide. Multiple bond descriptors reveal that isolated monovacancies undergo both in-plane reconstruction and out-of-plane distortion, which together shape their stability and electronic structure. For compact vacancy aggregates, we show that bond reconstruction acts as a key stabilization mechanism. In the carbon monovacancy, reconstruction qualitatively alters the electronic structure, yielding a singlet ground state and limiting its suitability as an optically addressable spin defect. In contrast, a highly stable aggregate composed of three carbon vacancies surrounding a silicon vacancy emerges as a promising infrared color-center candidate, combining a triplet ground state with a favorable Debye-Waller factor of the emission. These results highlight the role of bond reconstruction in defining the quantum properties of vacancy defects in two-dimensional silicon carbide.
\end{abstract}

\begin{document}
\title{Bond reconstruction and vacancy clustering in monolayer silicon carbide from first principles}
\author{P\'eter Udvarhelyi}
\email{udvarhelyi.peter@nims.go.jp}
\affiliation{International Center for Young Scientists, National Institute for Materials Science, 1-1 Namiki, Tsukuba, Ibaraki 305-0044, Japan}
\maketitle

\section{Introduction}
Two-dimensional materials are promising platforms for defect-based quantum technologies~\cite{Liu_2019, Montblanch_2023}. Their surface proximity offers several advantages over bulk qubit hosts, such as enhanced sensitivity and improved spatial resolution in quantum sensing~\cite{Gottscholl_2021, Udvarhelyi_2023}, mechanical fabrication methods for scalable, deterministic defect creation~\cite{Xu_2021, Ahmed_2025, Luo_2025}, and engineered quantum properties through surface functionalization~\cite{Gavin_2025}. The established host materials for defect-based quantum applications are either bulk materials, e.g., diamond, silicon, and silicon carbide (SiC), or layered bulk materials, e.g., hexagonal boron nitride (hBN) and various transition metal dichalcogenides (TMDs)~\cite{Lin_2016, Lee_2022}. Recent experimental progress has demonstrated emitter and qubit defects in few-layer and monolayer materials~\cite{Tran_2016, Toth_2019, Durand_2023}. Monolayer SiC has emerged as a promising material platform for hosting quantum defects, originating from the covalently bonded Si-C bilayer motif characteristic of bulk hexagonal SiC polytypes. Despite its reduced dimensionality, the monolayer retains many favorable attributes of bulk SiC, including moderate spin-orbit coupling, a low nuclear-spin density, a wide bandgap, and strongly covalent $sp^2$-hybridized bonding. Together, these features provide an attractive combination of mechanical robustness and magnetic coherence, supporting the stable incorporation and operation of quantum defects.
Synthesizing monolayer SiC is inherently difficult as SiC lacks a layered bulk analogue. Nevertheless, recent advances in bottom-up synthesis have enabled the successful realization of monolayer SiC through multiple routes, including epitaxial growth on ultrathin films~\cite{Polley_2023} and the formation of freestanding monolayer SiC within graphene nanopores~\cite{Da_2024}, providing experimentally viable platforms for exploring its intrinsic properties and defect physics. Additionally, theoretical calculations confirmed the stability of the atomically flat monolayer and showed reasonable agreement with experimental structures~\cite{Bekaroglu_2010, Shi_2015, Chabi_2020}.

Beyond quantum-defect applications, monolayer SiC has also attracted broader interest as a prototypical two-dimensional SiC-derived material platform. Its honeycomb lattice preserves the fundamental Si-C bonding motif of hexagonal SiC polytypes while eliminating the dangling bonds associated with exposed SiC surfaces, making it a useful model system for studying defect physics in reduced dimensions. Recent theoretical studies have further explored van der Waals heterostructures and chemically functionalized derivatives based on monolayer SiC, highlighting the versatility of this material beyond quantum-technology applications~\cite{Khakpour_2026, Hashemi_2026}. 

Motivated by these broader opportunities, intrinsic defects in monolayer SiC have been the subject of extensive first-principles investigations~\cite{Bekaroglu_2010, He_2010, Kuzubov_2012, Hassanzada_2020, Mahendiran_2022, Huang_2023, Singh_2023, Oktavina_2024, Mohseni_2024, Huang_2025, Haque_2025, Bartashevich_2025, Patra_2025}. Recent computational studies have focused particularly on monovacancy and divacancy defects, proposing optically active centers and spin-qubit candidates~\cite{Mohseni_2024, Huang_2025, Haque_2025}. Several of these works describe vacancy configurations that retain high-symmetry geometries and dangling-bond states. In contrast, Oktavina {\it et al.}~\cite{Oktavina_2024} reported symmetry lowering and bond reconstruction in monovacancy defects, although their structural search was restricted to in-plane atomic displacements. More recently, Bartashevich {\it et al.}~\cite{Bartashevich_2025} identified a nonplanar reconstructed carbon vacancy featuring Si-Si bond formation and a quenched magnetic moment. Together, these studies indicate that bond reconstruction is an important aspect of vacancy physics in monolayer SiC.

Despite these advances, the role of bond reconstruction in determining the stability, electronic structure, and optical properties of vacancy defects remains poorly understood. In particular, it is unclear how reconstruction influences optically active defect states and whether the energetic driving force for bond formation extends beyond isolated vacancies to larger vacancy aggregates. Since bond reconstruction can substantially lower the energy of defect complexes through the formation of additional reconstructed bonds, it may also govern vacancy aggregation pathways and the resulting family of optically active defects in monolayer SiC.

In this work, we computationally investigate symmetry-breaking distortions in monovacancy defects to establish a consistent description of their bonding characteristics and relaxed geometries. We find that dangling bond reconstruction plays a central role in driving these distortions and stabilizing the structures. In addition, we systematically study vacancy aggregation and reveal that bond reconstruction favors the formation of particular compact, high-symmetry clusters in monolayer SiC.

\section{Methods}

We performed density functional theory calculations within the plane‑wave formalism as implemented in the VASP package~\cite{VASP1, VASP2, VASP3, VASP4}. A kinetic‑energy cutoff of 500 eV was applied for the plane‑wave basis. Core electrons were treated using the projector‑augmented‑wave (PAW) method~\cite{PAW1, PAW2}. The model for the pristine SiC monolayer is a flat, graphene-like, hexagonal honeycomb lattice (P$\bar{\mathrm{6}}$m2 space group). The primitive unit cell contains one Si and one C atom. The lattice constant was optimized using the HSE06 functional~\cite{HSE} and a $12\times12\times1$ k-point mesh, yielding an equilibrium lattice constant of {3.072~\AA}. All defect calculations were subsequently performed using supercells constructed from this optimized structure. A periodic $10\times10$ supercell with 20~\AA~of vacuum separation between the layers was employed, which allows Brillouin-zone sampling to be restricted to the $\Gamma$-point. Reorientation reaction pathways were computed using the climbing-image nudged elastic band (CI-NEB) method~\cite{NEB, cNEB}. For the search and characterization of compact vacancy clusters, phonon calculations, and NEB calculations, the Perdew-Burke-Ernzerhof (PBE) functional was used~\cite{PBE}. All other results were obtained using the Heyd-Scuseria-Ernzerhof (HSE06) hybrid functional~\cite{HSE}. 
Bonding characteristics were evaluated using the LOBSTER package~\cite{Dronskowski_1993, Nelson_2020}. Integrated projected Crystal Orbital Hamiltonian Population (IpCOHP) analyses were performed for selected atom-pair interactions in the relaxed structures~\cite{Deringer_2011}. In each calculation, the integration upper limit was the corresponding Fermi energy, and negative IpCOHP values indicate bonding interactions.
Total energies in charged systems were corrected using the method of Freysoldt {\it et al.}~\cite{Freysoldt_2009, Freysoldt_2018}. Excited states were calculated using the $\Delta$SCF method of constrained orbital occupations~\cite{Xiong_2025}. Structural optimization was performed until all atomic forces were reduced below 0.01 eV/{\AA}. 

The formation energy in the $q$ charge state of the complex defect, formed by $n$ different $i$ species, was calculated as
\begin{equation}
E_{\text{f}}^{q}=E_{\text{defect}}^{q}-E_{\text{host}}-\sum_{i}\mu_i n_i +qE_{\text{Fermi}}\text{,}
\end{equation}
where $\mu_i$ is the chemical potential derived from bulk silicon and diamond. The Si-rich and C-rich limits of the chemical potentials were obtained using the calculated formation enthalpy of 4H-SiC.
This choice follows the conventional defect-formation formalism, where the chemical potentials are constrained by the thermodynamically stable phases of the constituent elements. The formation enthalpy was calculated according to 
\begin{equation}
 \Delta H_{\mathrm f}=E_{\mathrm{SiC}}-\frac{E_{\mathrm{Si}}+E_{\mathrm C}}{2}\text{,} 
\end{equation}
 where all energies are given per atom, and $E_{\mathrm{Si}}$ and $E_{\mathrm C}$ correspond to bulk silicon and diamond, respectively. The calculated HSE06 formation enthalpies are $-0.31$ eV/atom for bulk 4H-SiC and $+0.51$ eV/atom for monolayer SiC. Thus, within the present computational approach, monolayer SiC is metastable with respect to bulk 4H-SiC and therefore does not define the equilibrium thermodynamic limits of the Si-C system. We therefore use bulk 4H-SiC to determine the allowed range of silicon and carbon chemical potentials. We note that adopting monolayer SiC itself as the reference phase, as done in several previous studies, alters the carbon chemical potential under Si-rich conditions and the silicon chemical potential under C-rich conditions. Consequently, the formation energy of $\mathrm{V_C}$ in the Si-rich limit and that of $\mathrm{V_{Si}}$ in the C-rich limit depend on the chosen reference phase, whereas the corresponding formation energies at the opposite growth limits remain unchanged.

The binding energies of complex defects were calculated as
\begin{equation}
E_{\text{b}}=E_{\text{product}}-\sum_i^{j} E_{\text{constituent}, i}+(j-1)E_{\text{host}}\text{,}
\end{equation}
where the energies of products and constituents are taken from supercell calculations. Thus, the last correction term eliminates double counting of the lattice atoms. Negative binding energy is associated with bound states in this convention. 

Transition dipole moments were calculated using the ground-state pseudo-wavefunctions of the Kohn-Sham levels involved in the transition, as implemented in the VaspBandUnfolding code~\cite{Zheng}.
Zero-field splitting (ZFS) originating from the electron spin (S) dipolar interaction for $S>1/2$ axially symmetric systems is described by the Hamiltonian
\begin{equation}
    \hat{H}_{\text{SS}}=D\left(\hat{S}_{z}^2-\frac{S\left(S+1\right)}{3}\right)+E\left(\hat{S}_{x}^2-\hat{S}_{y}^2\right)\text{,}
\end{equation}
where $D$ and $E$ are the axial and rhombic ZFS parameters, respectively.
The hyperfine interaction between the electron spin (S) of the defect and nearby nuclear spins (I) in the lattice is described by the diagonalized Hamiltonian
\begin{equation}   
 \hat{H}_{\text{IS}}=A_{xx}\hat{I}_{x}\hat{S}_{x}+A_{yy}\hat{I}_{y}\hat{S}_{y}+A_{zz}\hat{I}_{z}\hat{S}_{z}\text{.}
\end{equation}
Both the ZFS and hyperfine parameters are calculated using the VASP package.
Atomic structures and isosurfaces are visualized in the VESTA code~\cite{Momma_2011}.

\section{Results and discussion}

\subsection{Bond reconstruction in monovacancy defects}

\begin{figure*}[!ht]
\centering
\includegraphics[width=0.85\linewidth]{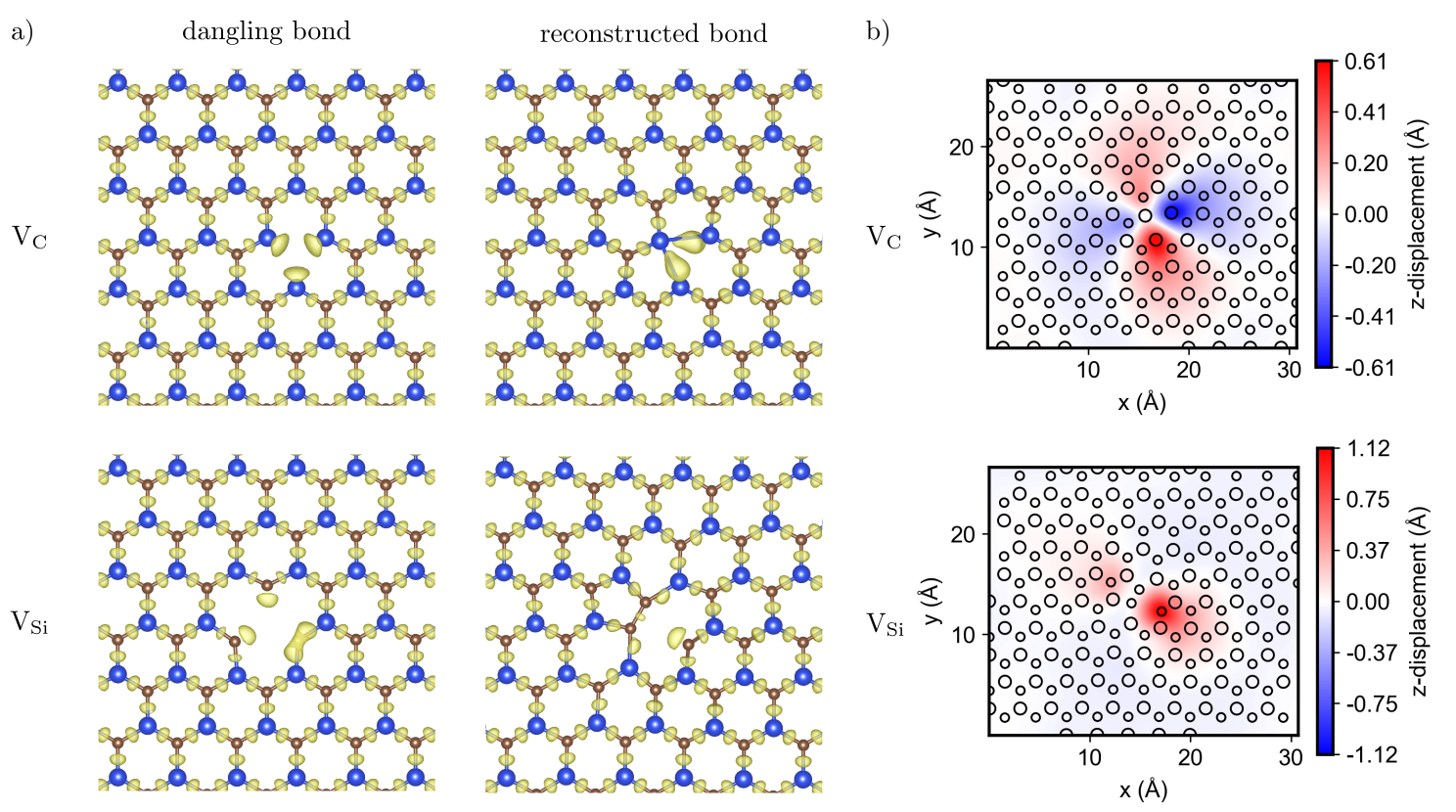}
\caption{Bond reconstruction and out-of-plane distortion in monovacancy defects in monolayer SiC. a) Electron localization function (ELF) at the 0.7 isovalue level (yellow) in the $\text{V}_\text{Si}$ and $\text{V}_\text{C}$ defects, comparing dangling and reconstructed bond calculations. The presence of nodal planes and the localization and direction of the lobes differentiate the two cases. Silicon and carbon atoms are illustrated as blue and brown balls, respectively. b) Accompanying out-of-plane distortion after bond reconstruction in $\text{V}_\text{Si}$ and $\text{V}_\text{C}$ defects, respectively. The displacement field is visualized as a heatmap projected onto the monolayer. Atomic sites of silicon and carbon are illustrated as larger and smaller circles, respectively.}
\label{fig:ELF}
\end{figure*}

\begin{figure}[!ht]
\centering
\includegraphics[width=\linewidth]{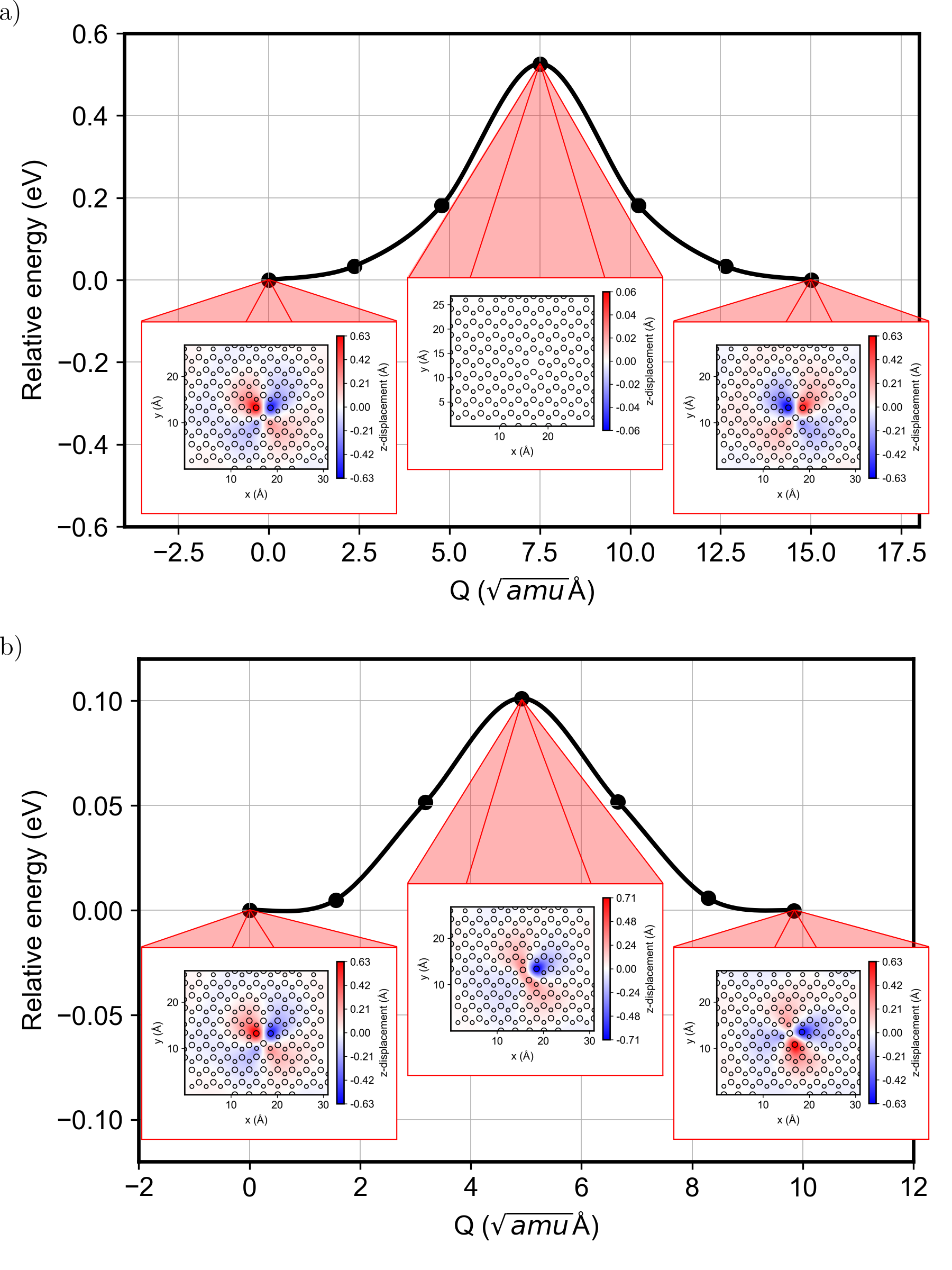}
\caption{Reorientation of chirality in the $\text{V}_\text{C}$ defect calculated using the CI-NEB method at the PBE functional level. a) Chirality swap reaction path through an in-plane barrier geometry. b) Chirality swap and rotation through an out-of-plane barrier geometry. Insets show the out-of-plane distortion heatmap projected on the SiC monolayer plane in the corresponding structures along the reaction path.}
\label{fig:NEB}
\end{figure}

\begin{figure*}[t]
\centering
\includegraphics[width=0.95\linewidth]{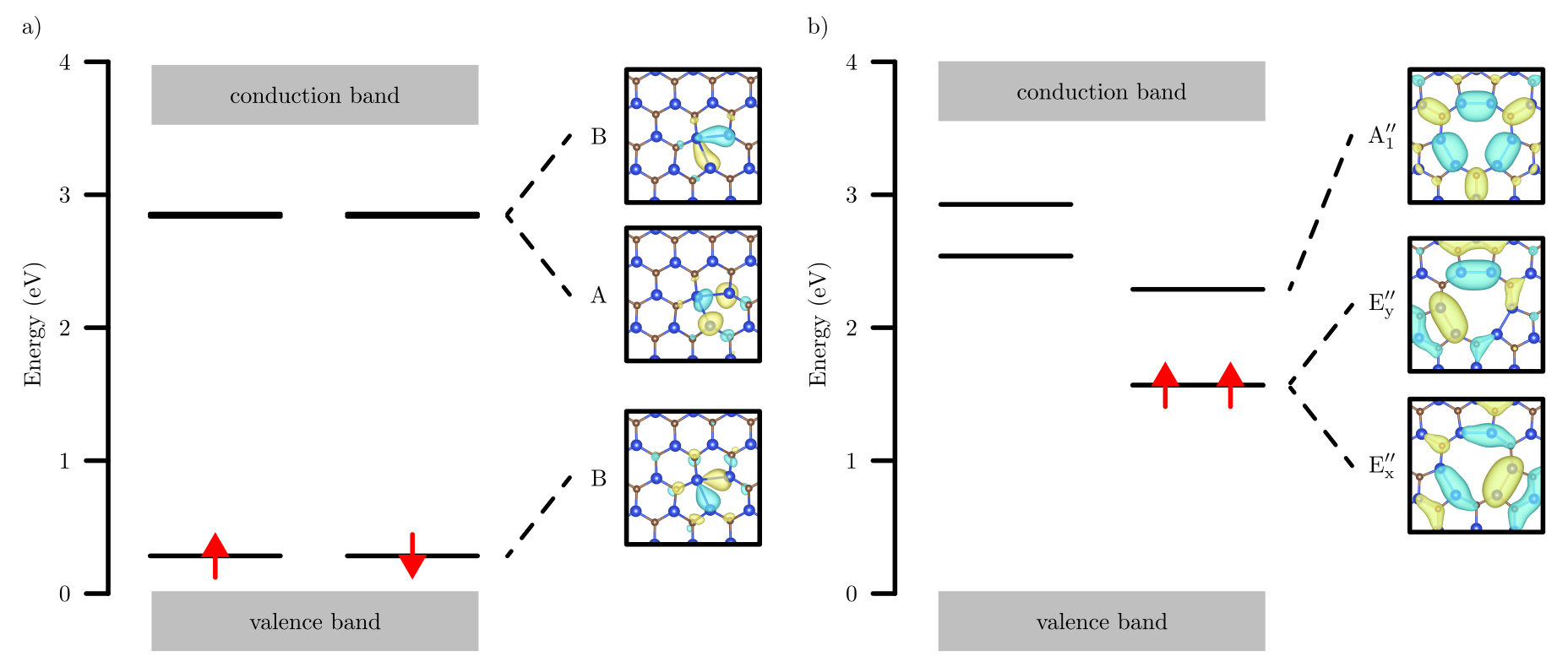}
\caption{Ground state Kohn-Sham level structures in spin-polarized HSE06 calculations in a) the $\text{V}_\text{C}$ and b) the $\text{V}_\text{Si}3\text{V}_\text{C}$ defects. Defect orbitals inside the bandgap are labelled according to the irreducible representations of a) $\text{C}_{2}$ and b) $\text{D}_{3\text{h}}$ point groups. Insets show the pseudo-wavefunction of the corresponding defect orbitals. Excited states are described via spin-conserving single-electron promotion to the unoccupied levels.}
\label{fig:levels}
\end{figure*}

Upon removing a single atom from the SiC monolayer, the vacancy retains the $\text{D}_{3\text{h}}$ point symmetry of the host lattice, enabling orbital degeneracy that favors high-spin states. Earlier studies identified a broad stability range for the neutral charge state under this symmetry and predicted a triplet ground state for both $\text{V}_\text{Si}$ and $\text{V}_\text{C}$ defects~\cite{Mohseni_2024}. We release this symmetry constraint in two steps, starting with in-plane reconstruction. For the neutral $\text{V}_\text{C}$ defect, we calculate a spin-singlet ground state in $\text{C}_{2\text{v}}$ symmetry and a reconstruction energy of 0.46 eV relative to the high-symmetry triplet state. The geometry shows a double reconstruction of the silicon dangling bonds, resulting in the 5-6-6-5 local pattern reported in Ref.~\onlinecite{Oktavina_2024}. Although previous calculations showed that the antisite form of the $\text{V}_\text{Si}$ defect possesses a significantly lower formation energy, our comparison focuses on the on-site structures. Within this restriction, in-plane relaxation yields a single reconstructed carbon bond, retaining a single carbon dangling bond. The resulting spin-triplet ground state lies 0.15 eV lower than the high-symmetry starting point. Lifting the planar-symmetry constraint in both defects reveals a spontaneous out-of-plane symmetry breaking, further decreasing their formation energies. For the $\text{V}_\text{C}$ defect, this result is consistent with the nonplanar reconstructed geometry previously reported by Bartashevich {\it et al.}~\cite{Bartashevich_2025}. The final structures of the neutral $\text{V}_\text{C}$ and $\text{V}_\text{Si}$ exhibit $\text{C}_{2}$ and $\text{C}_{1\text{h}}$ symmetries, respectively. Both defects possess a singlet spin ground state. However, the $\text{V}_\text{Si}$ ground-state wavefunction is symmetry-breaking, in line with previous reports~\cite{Mohseni_2024}. The total relaxation energies referenced to the corresponding structures in $\text{D}_{3\text{h}}$ symmetry are 1.03 eV and 0.31 eV, respectively. Because these values are more than double the reconstruction energy in the restricted planar models, this highlights that out-of-plane relaxations play a critical role in stabilizing the reconstructed structures. The largest out-of-plane displacements are 0.58~{\AA} and 1.05~{\AA} in the $\text{V}_\text{C}$ and $\text{V}_\text{Si}$ defects, respectively. The final formation energies of the neutral $\text{V}_\text{C}$ and $\text{V}_\text{Si}$ defects are 3.28 eV and 8.99 eV in Si-rich conditions, and 3.59 eV and 8.68 eV in C-rich conditions, respectively. Given this large energy difference, we focus on the $\text{V}_\text{C}$ defect in the further discussion.

\begin{table}[t]
\caption{Silicon bond reconstruction of vacancy defects in monolayer SiC. The analysis is referenced to the bond in bulk silicon. Calculations compare the shortest Si-Si distance and the Integrated projected Crystal Orbital Hamiltonian Population (IpCOHP). The point symmetries of the defects are listed in parentheses.}
\centering
\begin{tabular}{c c c}
system & Si-Si distance ({\AA}) & IpCOHP (eV) \\
\hline
bulk silicon reference
  & 2.35 & -5.3 \\

$\text{V}_\text{C}$ $(\text{D}_{3\text{h}})$
  & 2.99 & -1.3 \\

$\text{V}_\text{C}$ $(\text{C}_{2})$
  & 2.67 & -3.4 \\

$\text{V}_\text{Si}\text{V}_\text{C}$ $(\text{C}_{2\text{v}})$
  & 2.50 & -4.9 \\

$\text{V}_\text{Si}2\text{V}_\text{C}$ $(\text{C}_{2\text{v}})$
  & 2.60 & -4.3 \\

$\text{V}_\text{Si}3\text{V}_\text{C}$ $(\text{D}_{3\text{h}})$
  & 2.46 & -5.3
\end{tabular}\label{tab:IpCOHP}
\end{table}

The bond reconstruction is qualitatively analyzed by studying the topology of the Electron Localization Function (ELF)~\cite{ELF} plotted in Fig.~\ref{fig:ELF} a). We distinguish unperturbed covalent bonds and reconstructed bonds around the defects from dangling bonds by the position of the point attractors. While covalent-bond attractors are localized between atomic sites, dangling bonds exhibit attractors positioned between the vacancy site and its first neighbors. Comparing the basins around reconstructed bonds reveals weaker Si-Si bonding than that of the C-C reconstruction. IpCOHP calculations serve as a quantitative bonding characterization of the reconstructed $\text{V}_\text{C}$ defect. The results shown in Table~\ref{tab:IpCOHP} confirm a weak interaction in the constrained high-symmetry $\text{V}_\text{C}$ $(\text{D}_{3\text{h}})$ defect, consistent with dangling bonds, while bond reconstruction in the symmetry-breaking $\text{V}_\text{C}$ $(\text{C}_{2})$ defect results in a strong stabilization and a decreased interatomic distance, suggesting a weak covalent bond.

The comparison of the out-of-plane displacement pattern of the monovacancy defects is plotted in Fig.~\ref{fig:ELF} b), showing antisymmetric and symmetric displacements, consistent with $\text{C}_{2}$ and $\text{C}_{1\text{h}}$ point symmetry, for the $\text{V}_\text{C}$ and $\text{V}_\text{Si}$ defects, respectively. The former exhibits chirality, evidenced by the right-handed helicity of the three Si atoms around the vacancy site in this calculation. We analyze chirality swapping in the $\text{V}_\text{C}$ defect by calculating the reaction path in two interactions at the PBE level. Chirality swapping from left-handed to right-handed without rotation and together with a threefold rotation are shown in Fig~\ref{fig:NEB} a) and b), respectively. The reaction barrier in the former interaction corresponds to the in-plane-reconstructed defect with $\text{C}_{2\text{v}}$ symmetry. However, the reaction path involving rotation of the reconstruction pattern exhibits a much lower barrier energy corresponding to a structure with $\text{C}_{1\text{h}}$ symmetry. The calculated 0.1 eV reaction barrier suggests that chirality swapping and the reorientation of the bond-reconstruction pattern are dynamically averaged at elevated temperatures.

Bond reconstruction in the $\text{V}_\text{C}$ defect affects its calculated ground-state spin and orbital structure. As illustrated in Fig.~\ref{fig:levels} a), there is a much larger level separation between the occupied and unoccupied levels than was reported before in the $\text{D}_{3\text{h}}$ triplet ground state~\cite{Mohseni_2024}.
This qualitative change in the electronic structure revises the interpretation proposed in Ref.~\onlinecite{Mohseni_2024}, where the high-symmetry configuration of $\text{V}_\text{C}$ was identified as a potential optically addressable spin defect. In the present work, unrestricted structural relaxation stabilizes a reconstructed singlet ground state, eliminating the spin-triplet ground state required for such applications. Furthermore, the lowest-energy optical excitation of the reconstructed defect is found to be unstable with respect to structural relaxation, suggesting inefficient radiative emission and further limiting the suitability of isolated $\text{V}_\text{C}$ as a quantum-sensing or spin-photon-interface candidate.
 The lower point group symmetry of $\text{C}_{2}$ also affects the symmetry of the defect levels. They can be assigned to the two irreducible representations (irreps) under $\text{C}_{2}$ point symmetry, A and B. Although $\text{C}_{2}$ is a chiral point group, chirality is not represented by its irreps, nor does it violate their assignment. We find two bonding orbitals (B) and a single antibonding orbital (A) inside the bandgap. The unoccupied A and B orbitals are accidentally degenerate with a small energy splitting of 0.017 eV. Excitation to either unoccupied defect orbital is dipole-allowed by symmetry. However, the B orbitals are chiral counterparts: the occupied and unoccupied B orbitals exhibit right-handed and left-handed chirality, respectively. This difference manifests as a small orbital overlap, which arises from their localization on opposite sides of the monolayer plane. Conversely, the matching chirality of the A orbital and the occupied B orbital yields a strong overlap. This latter $\text{B}\rightarrow\text{A}$ transition yields a calculated transition dipole moment (TDM) of 4.3 D. Despite the favorable orbital overlap, the excited state remains dissociative in our calculations and decays to the ground state through a non-radiative process. This behaviour arises because the electron is promoted into an antibonding orbital, reducing the strength of the reconstructed bond and inducing a local structural expansion. Simultaneously, the energy difference between the bonding and antibonding orbitals decreases. Eventually, the promoted electron is stabilized at an energy lower than that of the unoccupied orbital, driving the system to the ground state and restoring the bonding configuration. These findings indicate that the optical activity of the $\text{V}_\text{C}$ defect can be compromised, as the efficient photon-absorption channel relaxes predominantly via non-radiative pathways, resulting in poor quantum efficiency. The definitive assessment of this prediction requires approaches that can reliably capture correlated multireference excited states and electron-phonon coupling, which lies outside the scope of this work. Motivated by the hindered optical activity found in the single-vacancy defect, we search for a vacancy aggregate with stability comparable to the $\text{V}_\text{C}$ defect as an alternative color-center candidate.

\subsection{Formation of compact vacancy clusters}

\begin{figure}[!ht]
\centering
\includegraphics[width=0.95\linewidth]{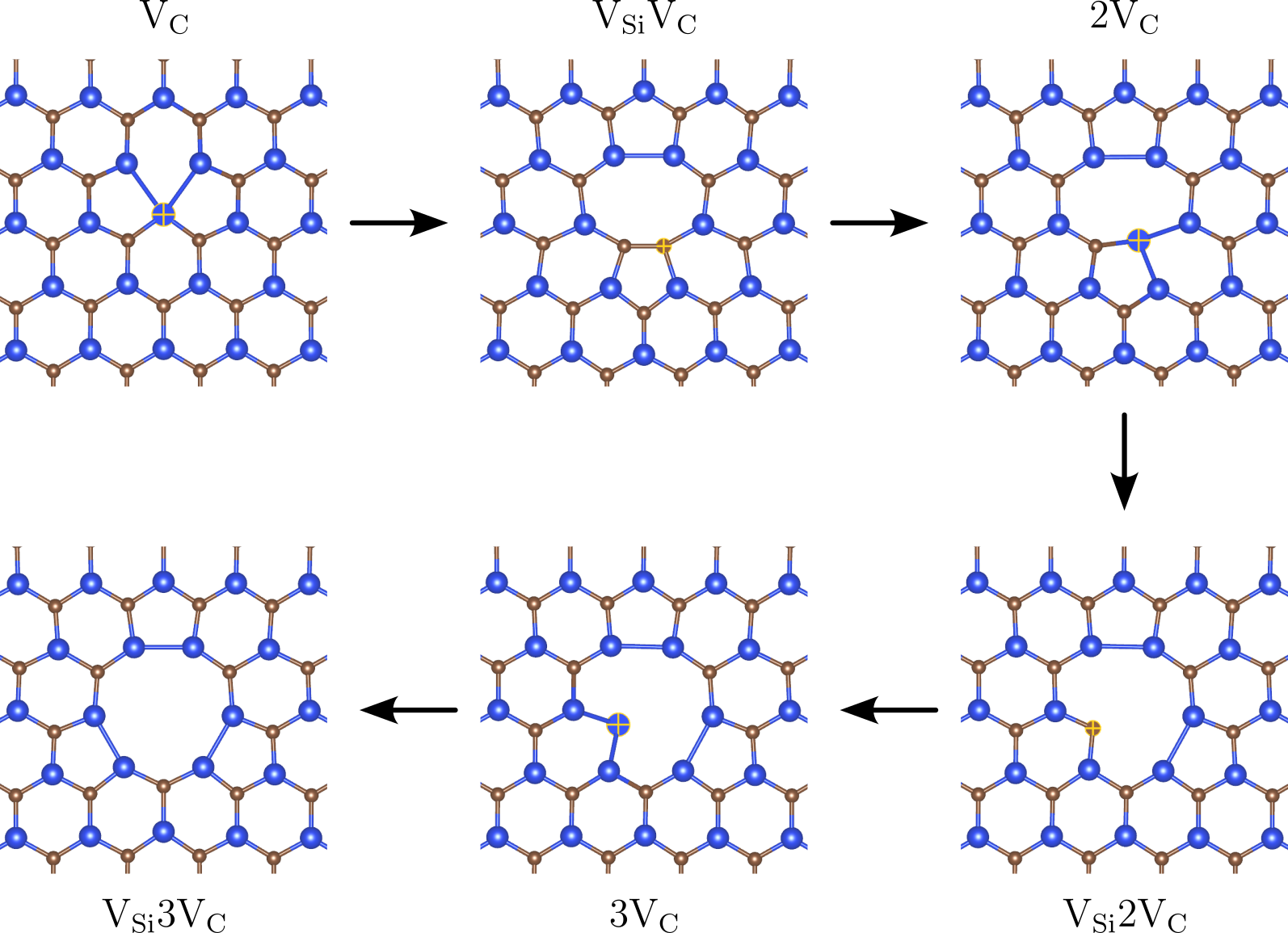}
\caption{Systematic construction of compact vacancy aggregate models. Atoms with dangling or frustrated bonds are considered as primary vacancy sites in the models, as shown in the $3\text{V}_\text{C}\rightarrow\text{V}_\text{Si}3\text{V}_\text{C}$ and $\text{V}_\text{C}\rightarrow\text{V}_\text{Si}\text{V}_\text{C}$ constructions, respectively. For the compact clusters that involve antisite defects, elements are swapped at the edge of the cluster or at reconstructed bonding sites, as illustrated in the $\text{V}_\text{Si}2\text{V}_\text{C}\rightarrow3\text{V}_\text{C}$ and $\text{V}_\text{Si}\text{V}_\text{C}\rightarrow2\text{V}_\text{C}$ constructions, respectively.}
\label{fig:method}
\end{figure}

\begin{figure*}[!ht]
\centering
\includegraphics[width=0.65\linewidth]{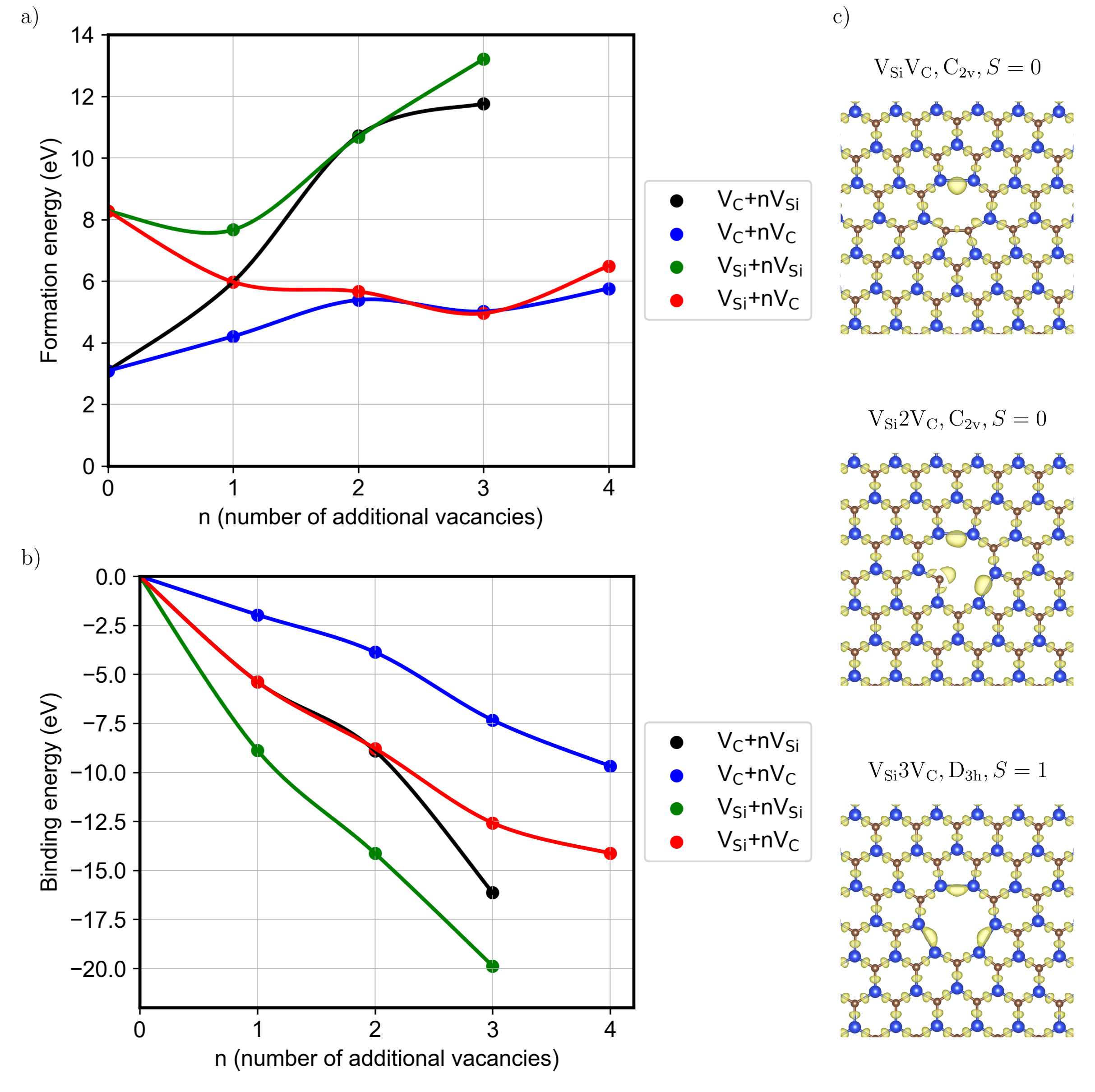}
\caption{Calculated structures of compact vacancy aggregates. a) and b) show the evolution of the formation and binding energies at the PBE level, respectively. Interpolated lines connect the calculation points to guide the eye. c) Geometry and ELF plots of the $\text{V}_\text{Si}+n\text{V}_\text{C}$ series. The ground-state point symmetry and spin are labelled for these structures.}
\label{fig:aggregation}
\end{figure*}

\begin{figure}[!ht]
\centering
\includegraphics[width=0.95\linewidth]{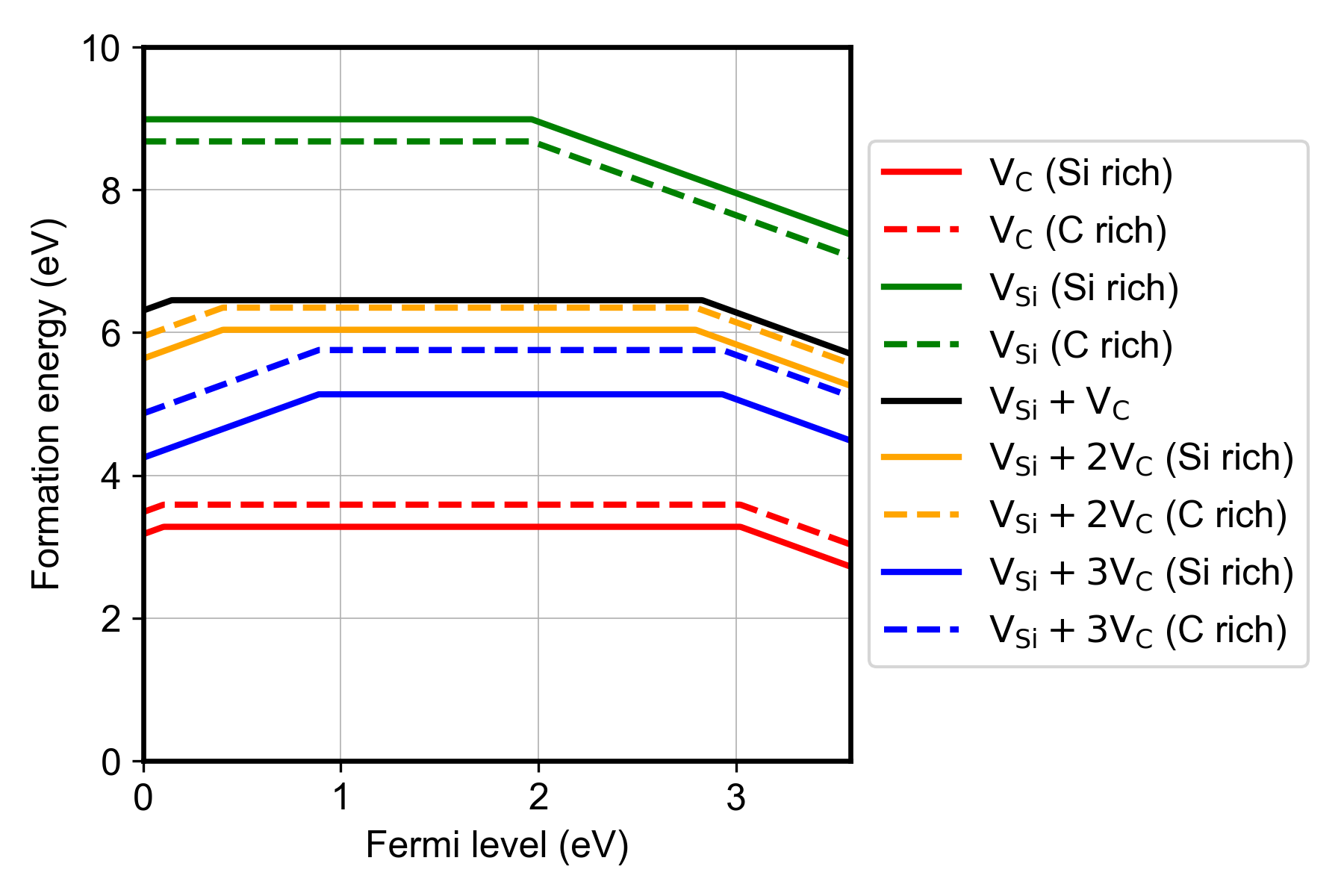}
\caption{Formation energy diagram of selected defects in HSE06 calculations plotted as a function of the Fermi level inside the bandgap referenced to the valence band maximum (VBM) energy.}
\label{fig:CTL}
\end{figure}

Computational insight into the aggregation of single-vacancy defects is important for assessing their stability and identifying vacancy clusters as potential candidates for quantum-technology applications. Moreover, bond reconstructions can play a similarly important role in determining their geometry and electronic structure, much like in monovacancy defects. 
Notably, the compact divacancy structures identified here differ qualitatively from previously reported high-symmetry dangling-bond configurations~\cite{Huang_2025}, emphasizing the strong energetic driving force for bond reconstruction in vacancy aggregates. The unreconstructed starting geometry is unstable in our calculations and consistently relaxes toward reconstructed structures in both singlet and triplet spin states. The reconstructed triplet state remains metastable in our calculations, but lies 0.87 eV above the reconstructed singlet ground state. Interestingly, the reconstructed triplet state exhibits even shorter reconstructed bond lengths than the singlet state, indicating that the singlet-triplet energy difference cannot be attributed solely to the presence of bond reconstruction. Instead, it arises from the different electronic occupations of the reconstructed defect states.
Consequently, we apply a restricted computational search for compact vacancy aggregates. We generate the initial configurations for the aggregation process by systematically removing or substituting active sites exhibiting dangling or reconstructed bonds around a monovacancy seed. Fig.~\ref{fig:method} outlines the model-construction workflow and shows that the restricted search reliably captures the relevant bond-reconstructed configurations. This justifies the assumption that our models sample the lowest-energy structures while avoiding the substantial computational overhead of an exhaustive high-throughput exploration. For relatively small cluster sizes, the aggregation process can be understood as vacancy clustering around the seed without losing generality. Consequently, we distinguish four evolution paths, consisting of the product of either $\text{V}_\text{Si}$ or $\text{V}_\text{C}$ seed and the aggregation of either $\text{V}_\text{Si}$ or $\text{V}_\text{C}$ defects around the seed. We label these processes accordingly, e.g., $\text{V}_\text{Si}+n\text{V}_\text{Si}$ labels the $\text{V}_\text{Si}$-seeded aggregation of $n$ addition $\text{V}_\text{Si}$ to form a compact cluster. We restrict our search to the neutral charge state, motivated by our goal of modeling an impurity-free, insulating material state. For selected candidates, the calculated stability range of the neutral charge state justifies this restriction. In this initial search, we use the PBE functional and calculate both singlet and triplet ground states. To identify particularly stable defect configurations, we analyze their formation ($E_\text{f}$) and binding energies ($E_\text{b}$), plotted in Fig.~\ref{fig:aggregation} a) and b), respectively. For larger vacancy clusters, $\text{V}_\text{C}$ aggregation is energetically favored over $\text{V}_\text{Si}$. Both $\text{V}_\text{Si}+n\text{V}_\text{C}$ and $\text{V}_\text{C}+n\text{V}_\text{C}$ show a local minium at $n=3$ in the formation energy plot. However, the $\text{V}_\text{Si}+n\text{V}_\text{C}$ series exhibits substantially stronger binding, with the binding energy of the $\text{V}_\text{Si}3\text{V}_\text{C}$ defect being approximately 5.2 eV lower than that of the corresponding $\text{V}_\text{C}3\text{V}_\text{C}$ structure. This indicates a significantly stronger thermodynamic driving force for aggregation in the former defect family. Upon relaxing the candidate structures, the preliminary search reveals that the $\text{V}_\text{Si}+n\text{V}_\text{C}$ sequence represents the most thermodynamically favorable aggregation route. This finding is particularly relevant for experimental scenarios involving elevated-temperature defect creation, where vacancy cluster formation is governed primarily by equilibrium energetics rather than kinetic constraints. Accordingly, we focus on this aggregation pathway in the following analysis. Fig.~\ref{fig:CTL} presents the complete formation-energy diagram, including the monovacancy defects as reference points. The results show a wide range of stability for the neutral charge state in all the calculated defects. This validates our {\it a priori} constraint to neutral systems studied in the preliminary search. The calculated defect energetics are sensitive to several methodological choices, including the treatment of structural relaxation, supercell size, exchange-correlation functional, chemical-potential reference, and finite-size charge corrections. These differences should be considered when comparing the present results with previous theoretical studies.

The calculated structures along the $\text{V}_\text{Si}+n\text{V}_\text{C}$ aggregation sequence are illustrated in Fig.~\ref{fig:aggregation} c). We identify reconstructed and dangling bonds via ELF topology analysis. All dangling bonds are reconstructed in the  $\text{V}_\text{Si}\text{V}_\text{C}$ and $\text{V}_\text{Si}3\text{V}_\text{C}$ defects, while a single dangling bond is retained in the $\text{V}_\text{Si}2\text{V}_\text{C}$ structure. The reconstruction in these defects preserves their original symmetry and planar structure. Table~\ref{tab:IpCOHP} shows the quantitative analysis of silicon-bond reconstruction. The calculated bond length and IpCOHP values reveal a weaker bond reconstruction in the $\text{V}_\text{Si}2\text{V}_\text{C}$ defect. In the $\text{V}_\text{Si}3\text{V}_\text{C}$ defect, the reconstructed silicon bonds are almost fully restored, showing bonding characteristics and bond lengths similar to bulk silicon. Notably, this defect is one of the unique vacancy aggregates that preserve the symmetry of the pristine monolayer and exhibit remarkable thermodynamic stability. The following section investigates its quantum properties in the context of quantum emitter applications.

\subsection{Quantum properties of the $\text{V}_\text{Si}3\text{V}_\text{C}$ defect}

\begin{table*}[t]
\caption{Calculated hyperfine coupling parameters in the $\text{V}_\text{Si}3\text{V}_\text{C}$ defect in monolayer SiC. Symmetrically equivalent atomic sites are grouped according to the $\text{D}_{3\text{h}}$ symmetry of the complex and labeled by their coordination shell and atomic species (e.g., 2n Si denotes the second-neighbor Si atoms around the vacancy cluster). Each label uniquely identifies a set of equivalent positions through their multiplicity and distance from the vacancy center. The eigenvalues of the total hyperfine tensor ($\mathbf{A}$) include the core contributions.}
\centering
\begin{tabular}{c c c c c c}
label & multiplicity & distance ({\AA}) & $A_{xx}$ (MHz) & $A_{yy}$ (MHz) & $A_{zz}$ (MHz) \\
\hline
1n Si & 6 & 2.89
  & -6.88 & -4.84 & -47.38 \\
2n Si & 3 & 5.06
  & -0.36 & -0.19 & -1.49 \\
3n Si & 3 & 5.34
  & -2.76 & -2.69 & -13.65 \\
3n Si & 6 & 6.07
  & -0.39 & -0.31 & -3.28 \\
5n Si & 6 & 8.13
  & -0.64 & -0.56 & -2.55 \\
1n C & 3 & 3.17
  & -4.73 & -4.49 & -7.34 \\
2n C & 6 & 4.63
  & -2.57 & -1.31 & -2.82
\end{tabular}\label{tab:hyper}
\end{table*}

\begin{figure}[t]
\centering
\includegraphics[width=0.95\linewidth]{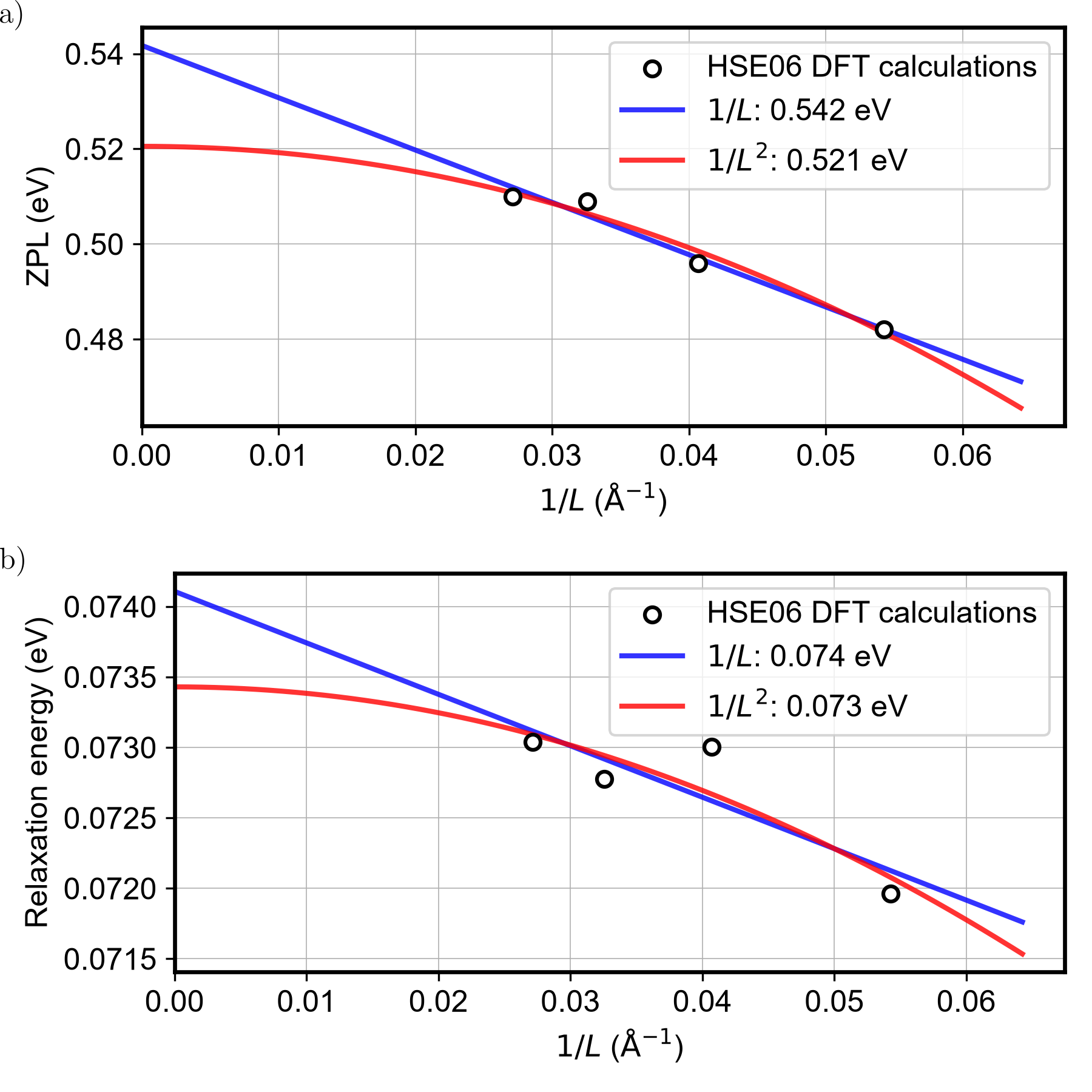}
\caption{Finite-size scaling of the calculated a) ZPL energy and b) structural relaxation energy in the excited state as a function of the inverse in-plane supercell size, $1/L$, where $L$ is the lattice parameter of the hexagonal supercell. The blue line shows a fit assuming a leading $1/L$ finite-size correction, $f(L)=a/L+b$, while the red line corresponds to a $1/L^2$ correction, $f(L)=a/L^2+b$.}
\label{fig:scaling}
\end{figure}

Our search for energetically favorable compact vacancy clusters resulted in the $\text{V}_\text{Si}3\text{V}_\text{C}$ defect combining a local formation energy minimum in the vacancy aggregation process and a relatively large binding energy, making it a candidate structure for a stable defect complex in the monolayer SiC. Its $S=1$ ground-state spin makes it a qubit defect, and its unpaired electron spin allows for zero-field splitting (ZFS) and hyperfine interactions. Our calculations yield a relatively small ZFS D parameter value of 214 MHz, consistent with the large extent of the four-vacancy complex. The spatial extension of the spin density reduces the dipolar electron-spin interaction, naturally leading to a weak ZFS. Table~\ref{tab:hyper} summarizes the calculated hyperfine interaction parameters. The negative values associated with the Si atom arise from its negative gyromagnetic ratio. Thus, positive spin density creates a negative interaction parameter. Conversely, C atoms interact with the negatively polarized regions of the spin density. Because the defect retains $\text{D}_{3\text{h}}$ symmetry, multiple atomic sites are symmetry-equivalent. The corresponding multiplicities are listed in the table.  We suggest the larger hyperfine values of the first-nearest-neighbor Si atoms (1n Si) and the closest third-nearest-neighbor Si atoms (3n Si) as characteristic fingerprints for experimental defect identification. Furthermore, our calculations reveal a great number of atomic sites in the close vicinity of the defect possessing hyperfine coupling parameters in the 1-10 MHz range, providing key information for the quantum memory application of the $\text{V}_\text{Si}3\text{V}_\text{C}$ defect. The unpaired electrons, together with their strongly localized hyperfine interaction on the first-neighbor Si atoms around the vacancy, indicate the existence of localized defect states.

\begin{figure}[t]
\centering
\includegraphics[width=\linewidth]{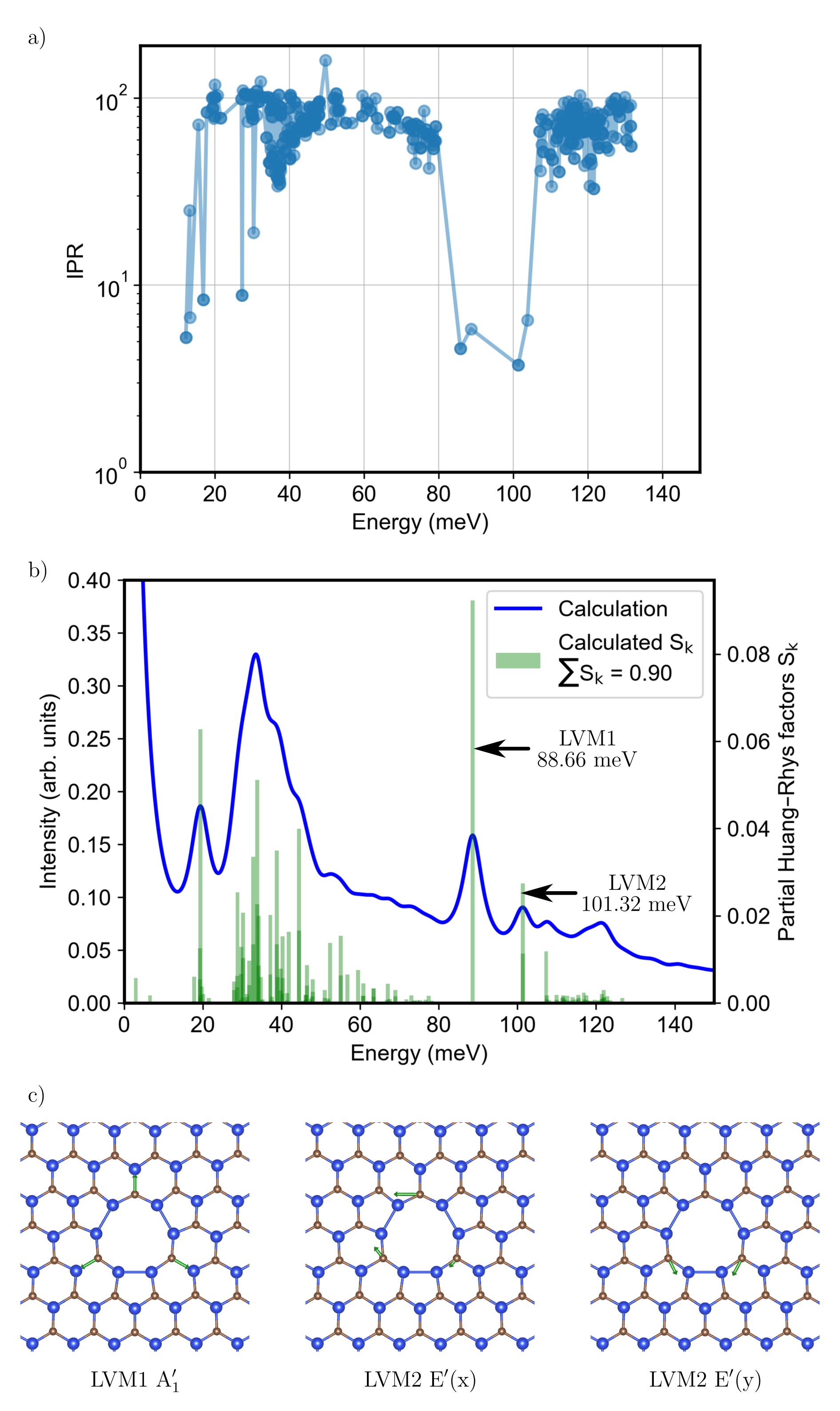}
\caption{Phonon calculations in the $\text{V}_\text{Si}3\text{V}_\text{C}$ defect at the PBE level. a) Calculated Inverse Participation Ratio (IPR) of the phonon modes in the ground state. b) Simulated phonon sideband of the Jahn-Teller active optical transition and the corresponding partial Huang-Rhys factors ($S_k$). Two optically active localized vibration modes (LVM) are identified in the sideband. c) Visualization of the displacement vectors in LVM1 and LVM2 localized modes.}
\label{fig:phonon}
\end{figure}

The calculated Kohn-Sham levels of the $\text{V}_\text{Si}3\text{V}_\text{C}$ defect inside the bandgap are illustrated in Fig.~\ref{fig:levels} b). The defect exhibits two localized levels, $E^{\prime\prime}$ and $A^{\prime\prime}_{1}$. The half-filled, doubly degenerate orbital results in the $S=1$ ground state. The first excited state of the defect is reached by promoting an electron to the unoccupied $A^{\prime\prime}_{1}$ level. This transition is electric-dipole allowed, with a calculated TDM of 12.5 D. The excited state exhibits Jahn-Teller (JT) instability and relaxes to a structure with $\text{C}_{2\text{v}}$ symmetry. To obtain the ZPL energy, we perform a finite-size correction by fitting the calculated ZPL values as a function of $1/L$ and $1/L^2$, where $L$ is the supercell lattice parameter, as shown in Fig.~\ref{fig:scaling} a). Both functional forms describe the available data with comparable accuracy. Given the neutral nature of the defect and the absence of long-range Coulomb interactions between periodic images, the $1/L^2$ scaling was adopted as the preferred extrapolation scheme. This yields a converged ZPL energy of 0.521 eV. The difference between the $1/L$ and $1/L^2$ extrapolations (21 meV) is taken as an estimate of the residual finite-size uncertainty. The resulting corrected ZPL energy of $0.52\pm0.02~\mathrm{eV}$ in the static JT limit lies in the near-infrared range. A similar finite-size extrapolation procedure was applied to the excited-state relaxation energy, as shown in Fig.~\ref{fig:scaling} b). It yields a converged value of $0.073\pm0.001~\mathrm{eV}$, where the uncertainty was estimated from the difference between the $1/L$1 and $1/L^2$ scaling models, as described above. The finite-size correction is substantially smaller than the typical numerical uncertainty associated with the DFT structural relaxation. This indicates that finite-size effects have a negligible impact on the calculated atomic relaxations for the supercells considered in this work.

Lastly, we evaluate phonon interactions in the $\text{V}_\text{Si}3\text{V}_\text{C}$ defect using PBE-level calculations. Inverse participation ratio (IPR) as a function of the phonon modes is shown in Fig.~\ref{fig:phonon} a). The logarithmic plot clearly separates bulk modes from localized vibration modes (LVM). Fig.~\ref{fig:phonon} b) shows the phonon sideband associated with the optical transition in the defect. It is generated by projecting the HSE06 level excited-state relaxation normal mode onto the ground-state vibrational modes. The calculated $S_k$ partial Huang-Rhys factors correspond to the coupling strength of the mode $k$ to the optical transition. We identify two optically active LVMs at 88.66 meV and 101.32 meV with respect to the ZPL energy. 
The atomic displacements in both LVMs are mostly localized on the symmetry-equivalent first-neighbor carbon atoms around the vacancy site. The corresponding displacement vectors are shown in Fig.~\ref{fig:phonon} c). LVM1 corresponds to a non-degenerate symmetric stretching vibration transforming as the $A_{1}^{\prime}$ irreducible representation of the local $\text{D}_{3\text{h}}$ symmetry. The higher-energy LVM2 belongs to the doubly degenerate $E^{\prime}$ representation, which is spanned by orthogonal basis functions corresponding to $E^{\prime}(x)$ antisymmetric-stretch-like and $E^{\prime}(y)$ bend-like distortions of the defect structure.
These sharp features in the optical signal are well isolated from the bulk modes and can serve as characteristic fingerprints for the experimental identification of the defect. The total Huang-Rhys factor of 0.9 corresponds to a Debye-Waller factor of 0.41, consistent with a relatively weak electron-phonon coupling. This promotes optical coherence in the emitter and also implies suppressed phonon-assisted non-radiative decay, leading to bright emission.

\section{Conclusion}

Our density functional theory calculations reveal that bond reconstruction plays a decisive role in stabilizing vacancy defects in monolayer SiC and shaping their quantum properties. For single vacancies, we find that out-of-plane distortions contribute substantially to stabilization, complementing the expected in-plane relaxation. Topological analysis of the electron density together with IpCOHP bond descriptors confirms partial bond reconstruction in the $\text{V}_\text{C}$ defect and full reconstruction in the $\text{V}_\text{Si}3\text{V}_\text{C}$ complex, explaining their favorable formation and binding energetics. While $\text{V}_\text{C}$ exhibits a potentially hindered optical excitation loop, the $\text{V}_\text{Si}3\text{V}_\text{C}$ defect shows robust optical characteristics, positioning it as a promising color-center candidate with a triplet spin. Its strong ground-state hyperfine interaction parameters and distinct LVM signatures in the optical spectrum provide clear experimental fingerprints for identification. Together, these insights establish a solid foundation for future theoretical and experimental efforts to realize vacancy-based quantum emitters in two-dimensional SiC.

\section*{Data availability statement}
All data supporting the findings in this study are included in the article. Raw calculation data are available from the authors upon reasonable request.

\section*{Acknowledgments}
P.U. acknowledges support from the ICYS research fund at the National Institute for Materials Science. This work was achieved through the use of the Numerical Materials Simulator at the National Institute for Materials Science.


%

\end{document}